\documentclass[aps,12pt,final,notitlepage,oneside,onecolumn,%
nobibnotes,nofootinbib,superscriptaddress,noshowpacs]{revtex4}
\usepackage{epsfig}
\newcommand{\eq}{\begin{eqnarray}}
\newcommand{\en}{\end{eqnarray}}

\begin{document}

\title{Strong CP violation and 
the neutron electric dipole form factor}
\author{\firstname{Jan}~\surname{Kuckei}}
\email{kuckei@tphys.physik.uni-tuebingen.de}
\affiliation{Institut f\"ur Theoretische Physik, 
Universit\"at T\"ubingen, \\
Auf der Morgenstelle 14, D-72076 T\"ubingen, Germany}
\author{\firstname{Claudio}~\surname{Dib}} 
\email{cdib@fis.utfsm.cl}
\affiliation{Departamento de F\'\i sica, Universidad
T\'ecnica Federico Santa Mar\'\i a, \\
Casilla 110-V, Valpara\'\i so, Chile} 
\author{\firstname{Amand}~\surname{Faessler}}
\email{amand.faessler@uni-tuebingen.de}
\affiliation{Institut f\"ur Theoretische Physik, 
Universit\"at T\"ubingen, \\
Auf der Morgenstelle 14, D-72076 T\"ubingen, Germany}
\author{\firstname{Thomas}~\surname{Gutsche}} 
\email{thomas.gutsche@uni-tuebingen.de}
\affiliation{Institut f\"ur Theoretische Physik, 
Universit\"at T\"ubingen, \\
Auf der Morgenstelle 14, D-72076 T\"ubingen, Germany}
\author{\firstname{Sergey}~\surname{Kovalenko}}
\email{Sergey.Kovalenko@usm.cl}
\affiliation{Departamento de F\'\i sica, Universidad
T\'ecnica Federico Santa Mar\'\i a, \\
Casilla 110-V, Valpara\'\i so, Chile}
\author{\firstname{Valery}~\surname{Lyubovitskij}} 
\email{valeri.lyubovitskij@uni-tuebingen.de}
\affiliation{Institut f\"ur Theoretische Physik, 
Universit\"at T\"ubingen, \\
Auf der Morgenstelle 14, D-72076 T\"ubingen, Germany}
\author{\firstname{Pumsa}~\surname{Pumsa-ard}} 
\email{pumsa@tphys.physik.uni-tuebingen.de}
\affiliation{Institut f\"ur Theoretische Physik, 
Universit\"at T\"ubingen, \\
Auf der Morgenstelle 14, D-72076 T\"ubingen, Germany}

\date{\today}

\begin{abstract}
We calculate the neutron electric dipole form factor induced by
the CP violating $\theta$-term of QCD, within a perturbative
chiral quark model which includes pion and kaon clouds. On this
basis we derive the neutron electric dipole moment and the
electron-neutron Schiff moment. From the existing experimental
upper limits on the neutron electric dipole moment we extract
constraints on the $\theta$-parameter and compare our results with
other approaches. 
\end{abstract}

\maketitle

\section{Introduction}

The understanding of CP violation is one of the long-standing and
challenging problems of particle physics. So far the effects of CP
violation have only been observed in the $K$ and $B$ hadron
systems~\cite{Eidelman:2004wy} and  are in good agreement with the
predictions of the Standard Model~(SM) of electroweak
interactions. Among other CP odd observables,
a great deal of effort has been directed to the study of electric
dipole moments (EDM) of leptons, neutrons and neutral atoms. Both
SM and various non-SM sources of CP violation have been considered
(for recent reviews see e.g. 
Ref.~\cite{rev-EDM,Erler:2004cx,Pospelov:2005pr}).
These studies have been particularly stimulated by the expectation
of great improvements (2 to 4 orders of magnitude) in the
experimental sensitivities to EDMs in the next decade (for review
see Ref.~\cite{Erler:2004cx}).

As it is well known, there are two sources of CP-violation within
the SM: the complex phase $\delta_{CKM}$ of the
Cabibbo-Kobayashi-Maskawa (CKM) quark mixing matrix
in the weak interaction sector, and the $\theta$-term in the strong
interaction sector, which arises due to the non-trivial structure
of the QCD vacuum~\cite{Belavin:1975fg}-\cite{Crewther:1978zz}.
The complex phase of the CKM matrix provides a
consistent explanation of the observed CP odd effects in hadron
decays, while it gives an imperceptible contribution to the EDMs,
far below the sensitivity of present or foreseeable experiments.
Indeed, the CKM prediction for the neutron EDM, $d_n$, ranges from
$10^{-31}$ to $10^{-33}$ $e \cdot$cm~\cite{Shabalin:1982sg} while
the present experimental upper limit~\cite{Harris:1999jx} is 
\begin{eqnarray}\label{EDM-exp-1}
|d_n| < 0.63 \times 10^{-25} \  e \cdot{\rm cm} \,.
\end{eqnarray} 
On the other hand, CP violation induced by a $\theta$-term leads
to a sizable electric dipole moment for
the neutron~\cite{Baluni:1978rf,Crewther:1979pi} and may
significantly contribute to atomic EDMs, while it is insignificant
for CP violation in hadron decays. The non-observation of the
neutron EDM and the atomic EDMs  imposes a very strict upper bound
on the value of $\theta$, of the order of $10^{-10}$. This unnaturally 
small value of $\theta$, which is otherwise not restricted by theory,
is known as the {\it strong CP problem}. One elegant solution was
proposed by Peccei and Quinn~\cite{Peccei:1977hh}, which makes the
$\theta$-parameter vanish dynamically. However, the mechanism also
requires the appearance of a Goldstone boson, the axion, which
remains to be discovered.

Other important contributions to the atomic and neutron EDMs may
arise from possible physics beyond the SM. In particular, 
supersymmetric extensions of the SM offer additional mechanisms for 
CP-violation~\cite{Weinberg:1989dx}-\cite{Demir:2003js}
originating from complex phases in the soft SUSY breaking terms and
superpotential parameters (for a review see Ref.~\cite{Khalil:2002qp}).

In the calculations of atomic and neutron EDMs one faces the
problem of translating the effect of the CP violation introduced 
at the quark-gluon level to the processes at the hadronic or atomic level. 
This translation must resort to hadronic and nuclear models as well as 
to a careful treatment of the atomic electron wave functions in the case 
of atomic EDMs. Therefore, confidence of the estimates of the hadronic 
or atomic CP-violating observables depends on reliability of these models.

The problem of the neutron EDM has been studied within various
theoretical approaches: current algebra and chiral perturbation
theory (ChPT)~\cite{Crewther:1979pi,Pich:1991fq,Borasoy:2000pq},
chiral quark models~\cite{Musakhanov:1984qy}-\cite{McGovern:1992ix},
lattice QCD~\cite{Aoki:1989rx,Shintani:2005xg},
QCD sum rules~\cite{Khatsimovsky:1987fr,Pospelov:1999mv},
an approach based on solutions of Schwinger-Dyson and 
Bethe-Salpeter equations~\cite{Hecht:2001ry}, etc.

In the present paper we apply to this problem the perturbative 
chiral quark model (PCQM)~\cite{Lyubovitskij:2001nm} 
that is a development of chiral quark models with a perturbative
treatment of the pion cloud of 
nucleon~\cite{Tegen:1983gg}-\cite{Gutsche:1989vy}. 
As shown in Ref.~\cite{Lyubovitskij:2001nm}, the PCQM is successful 
in description of low-energy properties of light baryons 
such as the mass spectrum, the electromagnetic, axial and strong 
form factors, including the quantities which receive a nontrivial 
contribution from the cloud of pseudoscalar mesons: meson-baryon 
sigma-terms, strangeness content of the nucleon, etc. 
Compared to the models of Refs.~\cite{Tegen:1983gg}-\cite{Gutsche:1989vy} 
the PCQM contains several new features\footnote{For details see 
Ref.~\cite{Lyubovitskij:2001nm}}: 
(i) the SU(3) extension of chiral symmetry in order to include 
the kaon and eta-meson cloud contributions; 
(ii) consistent formulation of perturbation theory both at the quark 
and baryon levels on the basis of 
the renormalization techniques and taking into account excited quark
states in the meson loop diagrams; 
(iii) incorporated constraints from the chiral symmetry (low-energy theorems);
(iv) consistency with chiral perturbation theory for the case of the  chiral
expansion of the nucleon mass.

The purpose of the present work is to calculate within the PCQM not 
only the neutron EDM but also the neutron electric dipole moment 
form factor (EDFF) induced by the strong CP violating $\theta$-term.  
The neutron EDFF as function of the momentum transfer could be the next step 
in the experimental studies of the CP odd structure of the neutron, after
the measurement of its EDM. Also, it is known~\cite{Thomas:1994wi} that 
the atomic EDMs are sensitive to the nuclear Schiff moment, 
which depends on the neutron 
EDM square radius derived from the neutron EDFF.

This paper is organized as follows. In Sect.~\ref{s2} we give a
brief introduction to the PCQM. Sect.~\ref{s3} deals with the
calculation of the electric dipole form factor, the 
neutron EDM and the electron-neutron Schiff moment induced by 
strong CP violation. Here, we extract the constraints for the CP 
violating $\theta$-parameter from the existing experimental data
on the neutron EDM and compare our results to some other
approaches. Finally, we give a summary of our results and
conclusions.

\section{The Perturbative Chiral Quark Model}
\label{s2}

The basis of the perturbative chiral quark model 
(PCQM)~\cite{Lyubovitskij:2001nm} is an effective chiral Lagrangian
describing the valence quarks of baryons as relativistic fermions
moving in an external field (static potential) 
$V_{\rm eff}(r)=S(r)+\gamma^0 V(r)$ with $r=|\vec x|$ , which in the
SU(3)-flavor version are supplemented by a cloud of Goldstone
bosons $(\pi, K, \eta)$. Treating Goldstone fields as small
fluctuations around the three-quark core, the linearized effective
Lagrangian is written as:
\begin{eqnarray}\label{L_eff}
{\cal L}_{\rm eff}(x)
&=& \bar q(x) [i \not\!\partial - S(r) - \gamma^0 V(r)] q(x) +
\frac{1}{2} \sum\limits_{i=1}^8  [\partial_\mu \Phi_i(x)]^2
\nonumber \\
&+& {\cal L}_I^{str (1)}(x) + {\cal L}_{\chi SB}(x).
\end{eqnarray}
Here we defined
\begin{equation}\label{L_eff_1}
{\cal L}_I^{str(1)}(x) =
- \bar q(x) i\gamma^5 \frac{\hat\Phi(x)}{F} S(r) q(x) \,.
\end{equation}
The additional term ${\cal L}_{\chi SB}$ in Eq.~(\ref{L_eff}) contains
the mass contributions both for quarks and mesons, which explicitly break
chiral symmetry:
\begin{equation}
{\cal L}_{\chi SB}(x) = -\bar q (x) {\cal M} q(x)
- \frac{B}{2} Tr [\hat \Phi^2(x)  {\cal M} ] \,.
\end{equation}
Here, $\hat \Phi = \sum\limits_{i=1}^8 \Phi_i \lambda_i$
is the octet matrix of pseudoscalar mesons,
$F=88$ MeV is the pion decay constant in the chiral
limit, ${\cal M}={\rm diag}\{\hat m,\hat m,m_s\}$
is the mass matrix of current quarks (we restrict to the isospin
symmetry limit $m_u=m_d=\hat m$) and $B=-<0|\bar u u|0>/F^2$ is the quark
condensate constant.
We rely on the standard picture of chiral symmetry
breaking and for the masses of pseudoscalar
mesons we use the leading term in their chiral expansion
(i.e. linear in the current quark mass):
\begin{equation}
M_{\pi}^2=2 \hat m B, \hspace*{.5cm} M_{K}^2=(\hat m + m_s) B,
\hspace*{.5cm} M_{\eta}^2= \frac{2}{3} (\hat m + 2m_s) B  \,.
\end{equation}
In our analysis we use the following set of parameters:
\begin{equation}
\hat m = 7 \;{\rm MeV},\; m_s =25 \hat m,\;
B = M^2_{\pi^+}/ 2 \hat m = 1.4 \;{\rm GeV} \,.
\end{equation}
The meson masses satisfy the Gell-Mann-Oakes-Renner and
Gell-Mann-Okubo relations. In addition, the linearized effective
Lagrangian fulfills PCAC. The properties of baryons, which are
modeled as bound states of valence quarks surrounded by a meson
cloud, are then derived using perturbation theory. At zeroth
order, the unperturbed Lagrangian simply describes a nucleon as
three relativistic valence quarks which are confined by an
effective one-body static potential $V_{\rm eff}(r)$ in the Dirac
equation. We denote the unperturbed three-quark ground-state as
$|\phi_0 \rangle$, with the normalization $\langle \phi_0 | \phi_0
\rangle = 1$. We expand the quark field $q$ in the basis of
eigenstates generated by this potential as
\begin{eqnarray}\label{q-expand}
q(x) = \sum_{\alpha} b_\alpha u_\alpha(\vec{x})
\exp(-i{\cal E}_\alpha t)
\end{eqnarray}
where the quark wave functions $\{ u_\alpha \}$ in orbits $\alpha$
are the solutions of the Dirac equation including 
the potential $V_{\rm eff}(r)$. 
The expansion coefficients $b_\alpha$ are the corresponding single
quark annihilation operators. All calculations are performed at an
order of accuracy~$o(1/F^2,\hat{m},m_s)$. In the
calculation of matrix elements, we project the quark diagrams on
the respective baryon states. The baryon states are conventionally
set up by the product of ${\rm SU(6)}$ spin-flavor and ${\rm SU(3)_c}$ 
color wave functions, where the nonrelativistic single
quark spin wave function is replaced by the relativistic solution
$u_\alpha(\vec{x})$ of the Dirac equation. 

In our description of baryons we use an effective potential
$V_{\rm eff}(r) = S(r) + \gamma^{0}V(r)$ which
is given by a sum of a scalar potential $S(r)$ providing
confinement and the time component of a vector potential
$\gamma^{0}V(r)$. Obviously, other possible Lorenz structures
(e.g., pseudoscalar or axial) are excluded by symmetry principles.
It is known from lattice simulations that a scalar potential should
be a linearly rising one and the vector potential is thought to be
responsible for short-range fluctuations of the gluon field
configurations~\cite{Takahashi:2000te}. In our study we approximate
$V_{\rm eff}(r)$ by a relativistic harmonic oscillator potential
with a quadratic radial dependence~\cite{Lyubovitskij:2001nm}
\begin{eqnarray}\label{V_hop}
S(r) = M_1 + c_1 r^2\,, \hspace*{.5cm}
V(r) = M_2 + c_2 r^2\,.
\end{eqnarray}
The model potential defines unperturbed wave functions for the quarks,
which are subsequently used to calculate baryon properties.
This potential has no direct connection to the underlying physical
picture and is thought to serve as an approximation of a realistic
potential. Note, that this type of the
potential was extensively used in chiral potential
models~\cite{Tegen:1983gg}-\cite{Gutsche:1989vy}.
A positive feature of this potential is that most
of the calculations can be done analytically. As was shown in
Refs.~\cite{Tegen:1983gg}-\cite{Gutsche:1989vy} and
later on also checked in the PCQM~\cite{Lyubovitskij:2001nm},
this effective potential gives a reasonable description of
baryon properties and can be treated as a phenomenological
approximation of the long-range potential dictated by QCD.

The use of a variational {\it Gaussian ansatz} for the 
effective potential (\ref{V_hop}) gives 
the following solution for the ground state (for the excited 
quark states we proceed by analogy):  
\begin{eqnarray}
u_0(\vec{x}) \, = \, N \, \exp\biggl[-\frac{\vec{x}^{\, 2}}{2R^2}\biggr]
\, \left(
\begin{array}{c}
1\\ i \rho \, \frac{\vec{\sigma}\vec{x}}{R}\\
\end{array}
\right) \, \chi_s \, \chi_f\, \chi_c \, ,
\end{eqnarray}
where $N=[\pi^{3/2} R^3 (1+3\rho^2/2)]^{-1/2}$ is a normalization
constant; $\chi_s$, $\chi_f$, $\chi_c$ are the spin, flavor and
color quark wave functions, respectively. The parameter $\rho$,
setting the strength of the "small component", can be related to
the axial charge $g_A$ of the nucleon. In the leading order
(3-quark-core) approximation, this relation
is~\cite{Lyubovitskij:2001nm}
\begin{equation}
g_A=\frac{5}{3}\biggl(1 - \frac{2\rho^2}{1+\frac{3}{2}\rho^2}\biggr) \,.
\end{equation}
The parameters of the effective potential $V_{\rm eff}(r)$ can also be
expressed in terms of $\rho$ and $R$:
\begin{equation}
M_1 = \frac{1 \, - \, 3\rho^2}{2 \, \rho R}, \;
M_2 = {\cal E}_0 - \frac{1 \, + \, 3\rho^2}{2 \, \rho R} , \;
c_1 \equiv c_2 =  \frac{\rho}{2R^3} \,.
\end{equation}
Here, ${\cal E}_0$ is the single-quark ground-state energy. In our
calculations we use the value $g_A$=1.25. Therefore, we have only
one free parameter in the model. In our numerical
study, $R$ is varied in the region from 0.55 fm to 0.65 fm,
which is set and constrained by nucleon
phenomenology~\cite{Lyubovitskij:2001nm}. Such a variation 
of the parameter $R$ slightly changes the physical quantities 
up to 5\%~\cite{Lyubovitskij:2001nm}. 
In this paper we also test a sensitivity of the neutron EDM 
to a variation of $R$. 

The expectation value of an operator $\hat A$ is defined as
\begin{equation}\label{matrA}
\langle \hat A \rangle = ^B\!\!\langle\phi_0|\sum^{\infty}_{n=1}
\frac{i^n}{n!}\int d^4 x_1 \ldots \int d^4 x_n T[{\cal L}_I (x_1)
\ldots{\cal L}_I (x_n) \hat A] |\phi_0\rangle_c^B
\end{equation}
where ${\cal L}_I$ is the full interaction Lagrangian which
may contain both CP-even and CP-odd terms, as discussed below.
The superscript $"B"$ in Eq.~(\ref{matrA}) indicates that
the matrix elements are projected on the respective baryon states and
the subscript $"c"$ refers to contributions from connected graphs only.

For the evaluation of Eq.~(\ref{matrA}) we apply Wick's theorem
with the appropriate propagators for quarks and mesons. For the
quark field we use 
a vacuum Feynman propagator for a fermion in a binding potential. 
In the calculation of meson-quark loops we include only the ground 
state in the quark propagator, which leads to the following truncated form:
\begin{equation}
i G_q(x,y) = \langle 0 |T\{q(x)\bar q(y)\}| 0 \rangle \ \to \
\theta(x_0-y_0) u_0(\vec{x}) \bar u_0(\vec{y}) e^{-i{\cal E}_0
(x_0-y_0)}\,.
\end{equation} 
Note, that in our previous papers we estimated explicitly the 
contribution of the low-lying excited quark states in the 
quark propagator to the physical quantities. Their 
contribution is about $\sim 10 - 15$\% with respect to the ground state 
contribution. Therefore, a restriction of the quark propagator 
to the ground states is a reasonable  approximation. 
For completeness we also calculate the corrections 
to the neutron EDM due to the inclusion of excited quark states: 
the first $p$-states ($1p_{1/2}$ and $1p_{3/2}$ in the non-relativistic 
notation) and the second excited states ($1d_{3/2}, 1d_{5/2}$ and 
$2s_{1/2}$), i.e. we restrict to the low-lying excited states with 
energies smaller than the typical scale of
$\Lambda =  1$ GeV of low-energy approaches.

For the meson fields we use their free Feynman propagators:
\begin{equation}
i\Delta_{ij}(x-y)= \langle 0|T\{\Phi_i(x)\Phi_j(y)\}|0 \rangle
=\delta_{ij}
\int\frac{d^4k}{(2\pi)^4i}\frac{\exp[-ik(x-y)]}{M_\Phi^2-k^2-i\epsilon}.
\end{equation}

\section{Strong CP violation and the neutron EDM}
\label{s3}

In this section we study the electric dipole moment form factor (EDFF)
and the electric dipole moment (EDM) of the neutron induced by the
$\theta$-term, using our perturbative chiral quark model (PCQM).
We follow a chiral approach for the
treatment of the $\theta$-term~\cite{Crewther:1978zz},
\cite{Borasoy:2000pq}-\cite{McGovern:1992ix} and then apply
the PCQM for the calculation of hadronic matrix elements.

At the fundamental level, the QCD Lagrangian is:
\begin{equation}\label{QCD-Lag}
{\cal{L}}_{QCD} = - \frac{1}{2} \, {\rm tr} \big( G_{\mu \nu} \,
G^{\mu \nu} \big) + \bar q ( i {\cal{D}} \!\!\!\!\slash -
{\cal{M}} ) q + \frac{\theta}{16 \pi^{2}} {\rm tr}
\big(\widetilde{G}_{\mu\nu} G^{\mu\nu} \big)\,.
\end{equation}
Here, the last term is the CP-violating $\theta$-term that cannot
be eliminated due to the nontrivial topology of the QCD vacuum.
As usual, ${\cal{D}}_{\mu}$ is the covariant derivative,
${G}_{\mu\nu}$ is the gluon stress tensor (in SU(3) matrix notation)
and $\widetilde{G}_{\mu\nu} = (1/2)\epsilon_{\mu\nu\sigma\rho}
{G}^{\sigma \rho}$ is its dual tensor. As it is well known, doing
a chiral $U(1)$ transformation in flavor space, one can remove the
gluonic $\theta$-term from the Lagrangian (\ref{QCD-Lag}) and pass
it as a (CP-violating) complex phase to the quark mass operators.
For a small value of $\theta$ the CP violating term becomes (for details
see Refs.~\cite{Baluni:1978rf,Crewther:1979pi,Fujikawa:1979ay}):
\begin{equation}
{\cal L}_{CPV}^{str (0)} =  i \theta \bar m  \,  \bar{q}(x) \gamma_5 q(x)
\end{equation}
where $q(x)$ denotes a flavor triplet, and, consequently,
this term is a flavor-$SU(3)$ singlet.
The mass coefficient is:
\begin{equation}
\bar{m} = \frac{m_u m_d m_s}{m_u m_d + m_u m_s + m_d m_s},
\end{equation}
which would vanish if any of the flavors were massless. From this
term we construct the effective chiral Lagrangian. As usual,
we do this by introducing the chiral field
$e^{i \gamma_5 \hat\Phi/F}$ and expand it in powers of $\hat\Phi/F$:
\begin{eqnarray}\label{LCP}
{\cal L}_{CPV}^{str} &=& i \theta \bar m \bar q \gamma_5
\exp\left(i\gamma_5 \frac{\hat\Phi}{F} \right) q  =
 i \theta \bar{m} \bar q \gamma_5 q - \theta \bar{m} \bar q
\frac{\hat\Phi}{F} q + O(\hat\Phi^2) \,.
\end{eqnarray}
It turns out that the first term in the r.h.s. of
Eq.~(\ref{LCP}) does not contribute to the EDFF at one loop 
because the diagrams involving this vertex do not contain the spin-flip
structure $\vec\sigma_N \cdot \vec q$, the product of the neutron spin
operator $\vec\sigma_N$ and of the 3-momentum of the photon $\vec q$.
The leading contribution to the EDFF comes from the linear 
term of the expansion in Eq.~(\ref{LCP}). Neglecting higher order terms,
we adopt for our analysis the CP-violating interaction in the form:
\begin{equation}\label{L_CP1}
{\cal L}_{CPV}^{str (1)} =
- \theta\bar{m}\bar q \frac{\hat\Phi}{F} q \,.
\end{equation} 
To guarantee electromagnetic gauge invariance in non-covariant
approaches (see discussion in
Refs.~\cite{Miller:1997jr,Lyubovitskij:2001nm}) we have to work in
the Breit frame, where the momenta of
the initial neutron, final neutron and photon are defined
respectively as: 
\begin{equation}
p = \left( E, - \vec{q}/2 \right)\,, \; \;
p^\prime = \left(  E, \vec{q}/2 \right)\,, \; \;
q = p^\prime - p = \left( 0, \vec{q}\, \right).
\end{equation}
Here, $E = \sqrt{m_N^2 + \vec{q}^{\,2}/4}$ is the nucleon energy,
$m_N$ is the nucleon mass, and $q^2 \equiv - Q^2 = - \vec{q}^{\,2}$ is
the momentum transfer squared. The neutron EDFF, $D_n(Q^2)$, is defined 
in the standard way through the neutron matrix element of 
the electromagnetic current: 
\begin{eqnarray}\label{Dn-def}
\langle n(p^\prime)| \, J_{\mu}(0) \, |n(p)\rangle &=&
\bar{u}_n(p^{\prime}) \biggl[ \gamma_\mu \, F_n^1(Q^2)
\, + \, \frac{i}{2 m_N} \, \sigma_{\mu\nu} \, q^{\nu} \, F_n^2(Q^2) \\
&-& \sigma_{\mu\nu} \, \gamma_5 \, q^{\nu} \, \, D_n(Q^2)
\, + \, (\gamma_\mu \, q^2 \, - \, 2 \, m_N \, q_\mu) \, \gamma_5 \, A_n(Q^2)
\biggr] u_n(p) \,, \nonumber
\end{eqnarray}
where, in addition, $F_n^1(Q^2)$ and $F_n^2(Q^2)$ are the
well-known $CP$-even neutron electromagnetic form factors and
$A_n(Q^2)$ is the neutron anapole moment form factor. In our model
at one loop level the neutron EDFF, $D_n (Q^2)$, 
is given in the Breit frame by
\begin{eqnarray}\label{Dn}
\frac{E}{m_N} \,
\chi^\dagger_{_{N_{s^\prime}}} \,  
i \, \vec{\sigma}_N \, \cdot \, \vec{q} \,\,
\chi_{_{N_s}} \,\, D_n(Q^2)  &=& ^N\!\!\langle\phi_0|
\, \frac{i^2}{2!} \, \int  \, \delta(t) \,
d^4 x d^4 x_1 d^4 x_2 \, e^{-iqx} \\
&\times& T\left[ \, {\cal L}_I^{str (1)}(x_1) \,
{\cal L}_{CPV}^{str (1)}(x_2)
J_{0}(x) \, \right] \, |\phi_0\rangle_c^N
\nonumber
\end{eqnarray}
where $\chi_{N_s}$ and $\chi^\dagger_{N_{s^\prime}}$ are the
nucleon spin wave function (w.f.) in the initial and final state, 
$\vec{\sigma}_N$ is the nucleon spin operator, ${\cal L}_I^{str (1)}$ 
and ${\cal L}_{CPV}^{str(1)}$ are the linearized parts of the CP-even 
and CP-odd  Lagrangian parts [c.f.\ Eqs.~(\ref{L_eff_1}) and
(\ref{L_CP1}), respectively] describing the strong interactions of
pseudoscalar mesons and quarks; $J_{0}$ is the time-component of
the electromagnetic current of the charged pseudoscalar meson
fields:
\begin{equation}
J_{\mu} \, = \, e \, ( \pi^- \, i\partial_\mu\pi^+ \,
+ \, K^- \, i\partial_\mu K^+ ) \ + \ {\rm h.c.}
\end{equation}
The only non-vanishing diagram contributing 
to the strong CP-violating part of the neutron EDFF 
is the so-called meson-cloud diagram shown in 
Fig.1. This diagram has two contributions: from pion and from kaon
loops. The pion cloud effects have already been estimated in
different theoretical approaches~\cite{Crewther:1979pi,Borasoy:2000pq,Musakhanov:1984qy,Morgan:1986yy,McGovern:1992ix}.
In addition to the pion cloud
the kaon loop effects have been calculated in the framework of
Heavy Baryon Chiral Perturbation Theory HBChPT~\cite{Borasoy:2000pq}.
Here we include both pion and kaon cloud contributions.

The result for the strong CP-violating contribution to the neutron EDFF
is given by 
\begin{eqnarray} \label{Eq_1}
D_n^{str}(Q^2) &=& D_n^{str; \pi}(Q^2) + D_n^{str; K}(Q^2) \, \\[5mm]
D_n^{str; \Phi}(Q^2) &=& e \, c_\Phi^{str} \,
\frac{g_{\pi NN} \, \bar g_{\pi NN}}{2 E} \,
\int\frac{d^3k}{(2\pi)^3} \, \frac{F_{\pi NN}[(\vec{k} + \vec{q}\,)^2] \,
\bar F_{\pi NN}(\vec{k}^{\,2})}{w_\Phi(\vec{k}+\vec{q}\,) \,
w_\Phi(\vec{k})} \, \frac{2}{w_\Phi(\vec{k} + \vec{q} \,) \,+ \,
w_\Phi(\vec{k}\,)}
\end{eqnarray}
where  $\Phi = \pi$ or $K$, $w_\Phi(\vec{q}\,) = \sqrt{M_\Phi^2 +
\vec{q}^{\,\,2}}$ is the meson energy, and $c_\Phi^{str}$ is 
a SU(6) spin-flavor factor, which is $c_\pi^{str} = 1$ for the
pion-loop diagram and $c_K^{str} = 1/5$ for the kaon-loop diagram.
Notice that the expression for the neutron EDFF 
is written in terms of the strong CP-conserving $g_{\pi NN}$ and
CP-violating $\bar g_{\pi NN}$ pion-nucleon coupling constants,
which are basic quantities in the effective low-energy
pion-nucleon Lagrangian derived from QCD~\cite{Crewther:1978zz}.
The strong coupling constant $g_{\pi NN}$ satisfies the
Goldberger-Treiman relation:
\begin{equation}
g_{\pi NN} \, = \, g_A \, \frac{m_N}{F} ,
\end{equation}
where $g_A$ is the axial nucleon charge, whereas the CP-violating
coupling $\bar{g}_{\pi NN}$ is identical to the result derived in
the context of the chiral quark model of Ref.~\cite{McGovern:1992ix}:
\begin{equation}
\bar g_{\pi NN} \, = \theta \, \frac{\bar m}{F} \, \gamma  \,.
\end{equation} 
Here $\gamma$ is the isovector-scalar two-quark condensate in the
nucleon:
\begin{equation}
\langle N|\bar q \tau_3 q|N \rangle \, = \,
\gamma \, \bar u_N \tau_3 u_N \, .
\end{equation}
This factor $\gamma$ coincides with the so-called relativistic
reduction factor~\cite{Lyubovitskij:2001nm}. In the PCQM,
$\gamma \equiv 5/8$~\cite{Lyubovitskij:2001nm}. Finally, $F_{\pi NN}$
and $\bar F_{\pi NN}$ are the normalized strong CP-conserving and
CP-violating $\pi NN$ form factors, which regularize the divergent
loop integral.
In the PCQM these form factors are given by~\cite{Lyubovitskij:2001nm}
\begin{equation}
F_{\pi NN}(\vec{k}^{\,2}) \, = \, \exp( - \vec{k}^{\,2} R^2/4) \,
\biggl[ 1 + \frac{\vec{k}^{\,2} R^2}{8} \biggl( 1 - \frac{5}{3 g_A}
\biggr)\biggr] \,,
\end{equation}
and
\begin{equation}
\bar F_{\pi NN}(\vec{k}^{\,2}) \, = \, \exp( - \vec{k}^{\,2} R^2/4) \,
\biggl[ 1 + \frac{\vec{k}^{\,2} R^2}{4}
\frac{1 - 3g_A/5}{9g_A/5 - 1} \biggr]
\end{equation}
where $R = 0.6$ fm is the PCQM dimensional parameter defining
the quark wave function.
The numerical results for the strong CP-violating contributions from
the pion and kaon loops to the neutron EDFF 
as functions of the momentum transfer are displayed in Fig.2. 
Here, the following comment is in order. The lack of covariance 
in our model makes the reliability of the above results for 
the neutron EDMFF decreasing in the region of large momentum transfer, 
where the relativistic effects
become non-negligible. Their typical size is determined by the ratio 
$\vec{q}^{2}/\left(4m^{2}_{N}\right)$, where $\vec{q}$ is the
three-momentum transfer. Therefore, the relativistic corrections are 
expected to be linearly growing with $Q^2=\vec{q}^{2}$. However, 
in the region $Q^{2}=\vec{q}^{2}< 0.4\,\rm{GeV}^{2}$ they do not 
exceed $\sim 10\%$ and can be neglected. 

The neutron EDM, $d_n$, is defined as the value of the neutron EDFF 
at zero recoil:
\begin{equation}
\label{Eq_3}
d_n^{str; \Phi} \equiv D_n^{str; \Phi}(0) \, = \,
e \, c_\Phi^{str} \, \frac{g_{\pi NN} \bar g_{\pi NN}}{2 m_N}
\int\frac{d^3k}{(2\pi)^3} \, \frac{F_{\pi NN}(\vec{k}^{\,2}) \,
\bar F_{\pi NN}(\vec{k}^{\,2})}{w^3_\Phi(\vec{k}\,)} \,.
\end{equation}
Now, it is worth checking if our result for the pion-cloud
contribution to $d_n^{str; \pi}$ in Eq.~(\ref{Eq_3}) is consistent with
the model-independent prediction derived in Ref.~\cite{Crewther:1979pi}
for the leading term in the chiral expansion:
\begin{equation}\label{Eq_4}
\bar d_n^{str; \pi} \, = \, e \, \frac{g_{\pi NN} \,
\bar g_{\pi NN}}{4 \, \pi^2 \, m_N} \, {\rm log}\frac{m_N}{M_\pi} \,.
\end{equation}
To this end, we drop the normalized form factors $F_{\pi NN}$ and
$\bar F_{\pi NN}$ in Eq.~(\ref{Eq_3}) by substituting 
$F_{\pi NN}=\bar F_{\pi NN}=1$, and analyze this equation using
alternatively cutoff or dimensional regularizations. Both methods
of regularization give the same result, which also coincide with
Eq.~(\ref{Eq_4}). Therefore, our approach is consistent with QCD
in the local limit, when $F_{\pi NN} = \bar F_{\pi NN} =1$. The
nontrivial form factors $F_{\pi NN}$ and $\bar F_{\pi NN}$ provide
an ultraviolet convergence for the EDM. The leading contribution
of the kaon-cloud diagram is also proportional to the chiral
logarithm but contains a model-dependent coefficient $c_K^{str} = 1/5$:
\begin{equation}\label{Eq_4_K}
\bar d_n^{str; K} \, = \, c_K^{str} \, e \, \frac{g_{\pi NN} \,
\bar g_{\pi NN}}{4 \, \pi^2 \, m_N} \, {\rm log}\frac{m_N}{M_K} \,.
\end{equation}
We remark that the coefficient $c_K^{str}$ was calculated  
previously  in Heavy Baryon Chiral Perturbation
Theory~(HBChPT)~\cite{Borasoy:2000pq}, where it was  
expressed through the parameters of the chiral Lagrangian:
\begin{eqnarray}
c_K^{str} \, = \, \frac{D-F}{D+F} \, \frac{b_F-b_D}{b_F+b_D} .
\end{eqnarray}
Here $D$ and $F$ are the axial-vector couplings; $b_D$ and $b_F$ 
are the low-energy constants. Using the actual
values~\cite{Borasoy:2000pq} of $D = 0.80$ and $F = 0.46$ fixed
from a fit of semileptonic hyperon
decays, and $b_D = 0.079$ GeV$^{-1}$, $b_F = - 0.316$ GeV$^{-1}$
determined from the calculation of baryon masses and the $\pi N$
sigma-term up to fourth order in the chiral expansion, we deduce
the prediction of HBChPT of $c_K^{str} = 0.45$. This is more than
a factor two larger than the prediction of our model.

Our results for the neutron EDM induced by the $\theta$-term at one-loop 
approximation are:  

{\bf(a)} partial pion and kaon loop contributions
\begin{equation}\label{pi-K-part}
d_n^{str; \pi} = 1.37 \times 10^{-16} \times \theta  \ \ [e  \cdot
{\rm cm}] \,, \ \ \ d_n^{str; K} = 0.05 \times 10^{-16} \times
\theta  \ \  [e \cdot {\rm cm}] \,.
\end{equation}

{\bf (b)} the total pion and kaon loop contribution
\begin{equation}\label{tot1}
d_n^{str} = d_n^{str; \pi} + d_n^{str; K} =  1.42 \times 10^{-16}
\times \theta  \ \   [e \cdot {\rm cm}]\,.
\end{equation}
As seen from Eqs.~(\ref{pi-K-part}), the kaon contribution to the
neutron EDM is smaller than the pion contribution by a factor $\sim 28\,.$ 

Now, let us estimate the sensitivity of our results
for the neutron EDM on the model approximations:
(i) variation of the size parameter $R$ from 0.55 to 0.65 fm,
(ii) contribution of the excited quark states and
(iii) two-loop corrections.

(i) For the range of values $R = 0.6 \pm 0.05$ fm
the values of the quantities, defined in Eqs.~(\ref{pi-K-part}) and 
(\ref{tot1}), are varying within the following ranges:
\begin{eqnarray}
& &d_n^{str}  =  (1.42 \pm 0.1)\ \, \times 10^{-16}
\times \theta  \ \   [e \cdot {\rm cm}]\,, \ \
d_n^{str; \pi} =  (1.37 \pm 0.1) \times 10^{-16} 
\times \theta  \ \   [e \cdot {\rm cm}]\,, \\
&&d_n^{str;K} =  (0.05 \pm 0.01) \times 10^{-16}
\times \theta  \ \   [e \cdot {\rm cm}]\,. \nonumber
\end{eqnarray}

(ii) The contribution of the excited quark states to the quark propagator, 
neglected in the present study, according to our estimation is of
the order of 10\% both for the pion and kaon meson cloud diagrams.
This result is in agreement with the previous studies of
the electromagnetic nucleon form factors, the electromagnetic 
$N\to \Delta \gamma$ transition and the axial nucleon form 
factor~\cite{Lyubovitskij:2001nm}.

(iii) The two-loop corrections to the one-loop result can be roughly
estimated using a naive dimensional analysis~\cite{Manohar:1983md},  
from which it follows that their contribution is suppressed 
by a factor
\begin{eqnarray}
\epsilon = \biggl(\frac{\Lambda}{4 \pi F}\biggr)^2 =
\biggl(\frac{1}{4 \pi F R}\biggr)^2 \simeq 0.1 \, 
\end{eqnarray}
and, therefore, can be safely neglected at the level of accuracy 
adopted in the present analysis. 

The current experimental bound on the neutron EDM in
Eq.~(\ref{EDM-exp-1}) used in Eq. (\ref{tot1}) leads to  
the following upper limit for the QCD angle $\theta$:
\begin{equation}\label{theta-limit}
|\theta| < 4.4 \times 10^{-10} \,.
\end{equation}

The results of our model are in a reasonable agreement with
the predictions of other chiral approaches shown in Table I. 

It is worth noticing that our prediction for the kaon cloud contribution
$d_n^{str; K} = 0.05 \times 10^{-16} \times \theta  \ \  [e \cdot
{\rm cm}]$ is much smaller than the analogous result of
HBChPT~\cite{Borasoy:2000pq}: $d_n^{str; K} = 1.1 \times 10^{-16}
\times \theta  \ \  [e \cdot {\rm cm}]$.

Another CP-odd parameter which may have important implications for 
CP violation in atoms is the electron-neutron Schiff moment $S^{\prime}$, 
defined in terms of the neutron EDFF $D_n(Q^2)$ as~\cite{Thomas:1994wi}
\begin{eqnarray}\label{Shiff-def}
S^{\prime} = - \left[\frac{d D_n(Q^2)}{dQ^2}\right]_{Q^2=0} \,.
\end{eqnarray}
Using our present calculation for $D^{str}_n(Q^2)$ induced by 
strong CP-violation, we have
\begin{eqnarray}\label{Schiff-PCQM}
S^{\prime \, str; \Phi} &= &
e \, c_\Phi^{str} \, \frac{g_{\pi NN} \bar g_{\pi NN}}{2 m_N}
\int\frac{d^3k}{(2\pi)^3} \, \frac{F_{\pi NN}(\vec{k}^{\,2}) \,
\bar F_{\pi NN}(\vec{k}^{\,2})}{w^3_\Phi(\vec{k}\,)} \nonumber\\
&\times&\biggl[ \frac{3}{4 \, w^2_\Phi(\vec{k}\,)} -
 \frac{5}{6} \, \frac{\vec{k}^{\,2}}{w^4_\Phi(\vec{k}\,)}
+ \frac{1}{8 \, m_N^2} - \frac{M_\Phi^2}{w^2_\Phi(\vec{k}\,)} \,
\frac{F_{\pi NN}^{\prime}(\vec{k}^{\,2})}{F_{\pi NN}(\vec{k}^{\,2})}
- \frac{2}{3} \, \vec{k}^{\,2} \,
\frac{F_{\pi NN}^{\prime\prime}(\vec{k}^{\,2})}{F_{\pi NN}(\vec{k}^{\,2})}
\biggr]\,,
\end{eqnarray}
where $F_{\pi NN}^{\prime}$ and $F_{\pi NN}^{\prime\prime}$ are
the first and second derivatives of the $F_{\pi NN}$ form
factor with respect to $\vec{k}^{\,2}$.

Again, as in the case of the neutron EDM, we may check the 
consistency of our approach with ChPT at leading order in 
the chiral expansion. To this end we drop the normalized form factors 
$F_{\pi NN}$ and $\bar F_{\pi NN}$ in Eq.~(\ref{Schiff-PCQM}) 
substituting $F_{\pi NN}=\bar F_{\pi NN}=1$ and, therefore, 
$F_{\pi NN}^{\prime} = F_{\pi NN}^{\prime\prime} = 0$. Then,  
keeping the leading terms in the chiral expansion (which are 
ultraviolet-convergent) we get:
\begin{eqnarray}\label{Schiff-PCQM_LO}
\bar S^{\prime \, str; \Phi} =
e \, c_\Phi^{str} \, \frac{g_{\pi NN} \bar g_{\pi NN}}{2 m_N}
\int\frac{d^3k}{(2\pi)^3} \,
\frac{1}{w^5_\Phi(\vec{k}\,)} \, \biggl[ \frac{3}{4} -
 \frac{5}{6} \, \frac{\vec{k}^{\,2}}{w^2_\Phi(\vec{k}\,)} \biggr] .
\end{eqnarray}
Carrying out the integration in Eq.~(\ref{Schiff-PCQM_LO}) explicitly, 
we arrive at the expression:
\begin{eqnarray}\label{Schiff-PCQM_LO_2}
\bar S^{\prime \, str; \Phi} =
e \, c_\Phi^{str} \,
\frac{g_{\pi NN}\,\bar g_{\pi NN}}{48 \, \pi^2 \, m_N \, M_\Phi^2} \,.
\end{eqnarray}
In the case of the pion-cloud contribution ($c_\pi^{str} = 1$)
this result coincides with the leading order result of
ChPT~\cite{Thomas:1994wi,Hockings:2005cn}:
\begin{eqnarray}\label{Schiff_LO}
\bar S^{\prime} =
e  \, \frac{g_{\pi NN}\,\bar g_{\pi NN}}{48 \, \pi^2 \, m_N \, M_\pi^2}
\approx  4.1 \times 10^{-2} \times \theta  \times  [ e/{\rm GeV}^{3}] \,.
\end{eqnarray}
Here, in the numerical estimate of $\bar S^{\prime}$ we use the values 
$g_{\pi NN}^2/4\pi \approx 14$ and $\bar g_{\pi NN}=0.027 \,\times \theta$ 
from Ref.~\cite{Crewther:1979pi}. 

Finally, from Eq.~(\ref{Schiff-PCQM}) we obtain the following 
results for the electron-neutron Schiff moment $S^{\prime}$ in our model:

{\bf(a)} partial pion and kaon loop contributions
\begin{equation}\label{S_pi-K-part}
S^{\prime \, str; \pi} = 3.86 \times 10^{-2} \times \theta  \ \
[ e/{\rm GeV}^{3}] \,, \ \ \
S^{\prime \, str; K} = 0.05 \times 10^{-2} \times
\theta  \ \  [ e/{\rm GeV}^{3}] \,.
\end{equation}

{\bf (b)} the total pion and kaon loop contribution 
\begin{equation}\label{S_tot}
S^{\prime \, str} = S^{\prime \, str; \pi} + S^{\prime \, str; K} =
3.91 \times 10^{-2} \times \theta  \ \   [ e/{\rm GeV}^{3}]\,.
\end{equation}
The above results have been derived in one-loop approximation, 
with the quark propagator truncated to include only the ground state 
and for the size parameter value $R=0.6$ fm. 

As seen from Eq.~(\ref{S_pi-K-part}),
the kaon contribution to the electron-neutron Schiff moment is 
suppressed by a factor $\sim 10^{-2}$ compared to the pion contribution. 
Comparing our prediction for $S^{\prime}$ in Eq.~(\ref{S_tot}) with 
the leading order prediction of ChPT~\cite{Thomas:1994wi,Hockings:2005cn} 
shown in Eq.~(\ref{Schiff_LO}) we conclude that both results are 
numerically  rather close.

\section{Summary and Conclusions}
\label{summary}

In this work we applied the perturbative chiral quark model to the
calculation of the neutron electric dipole form factor induced by
a strong CP violating $\theta$-term. We have taken into account
both pion and kaon cloud contributions. From the existing
experimental constraints on the neutron electric dipole moment we
have derived an upper limit on the CP violating parameter
$\theta$, which is compatible with the corresponding limits of
other existing approaches. However, we have found that our
prediction for the kaon cloud contribution to the neutron EDM is
much smaller than the analogous result from Heavy Baryon Chiral
Perturbation Theory (HBChPT). We have also calculated the
electron-neutron Schiff moment, a quantity that can be used for
the calculations of the electron-nucleus Schiff moments on the
basis of specific nuclear models. We have found that our result is
numerically consistent with the leading order prediction from
chiral perturbation theory. 

In the present paper we have not considered the sources of 
the CP-violation beyond the SM, which, however, may have 
an important impact on the neutron electric dipole moment form factors.
The corresponding results in our model will be published elsewhere.

\begin{acknowledgments}
This work was supported in part by the DAAD under contract
415-ALECHILE/ALE-02/21672, by the FONDECYT projects 1030244,
1030254, 1030355, by the DFG under contracts FA67/25-3 and GRK683.
This research is also part of the EU Integrated Infrastructure
Initiative Hadronphysics project under contract number
RII3-CT-2004-506078 and President grant of Russia "Scientific
Schools"  No. 5103.2006.2. K.P. thanks the Development and Promotion
of Science and Technology Talent Project (DPST), Thailand for
financial support. C.D. and V.E.L. thank the Institute of
Theoretical Physics at the University of T\" ubingen, Germany and
the Departamento de F\'\i sica at the Universidad T\'ecnica
Federico Santa Mar\'\i a of Valpara\'\i so, Chile, respectively,
for their kind hospitality.
\end{acknowledgments}

\newpage

\vspace*{-3cm} 
\begin{figure}
\begin{center}
\epsfig{file=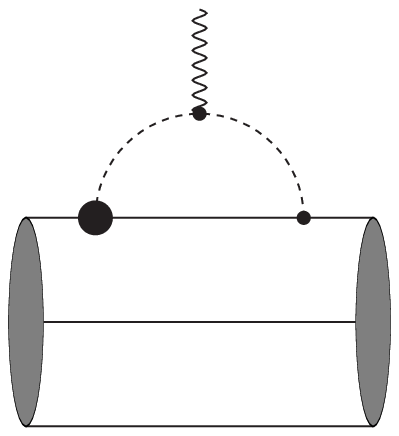, scale=0.8}
\end{center}
\vspace*{-6cm}
\caption{The strong CP-violating contribution to the neutron
electric dipole moment form factor: meson cloud diagram. The
vertex denoted by a black filled circle corresponds to the
insertion of the strong CP violating interaction of
Eq.~(\ref{L_CP1}).}
\end{figure}

\begin{figure}
\vspace*{1.5cm}
\begin{center}
\epsfig{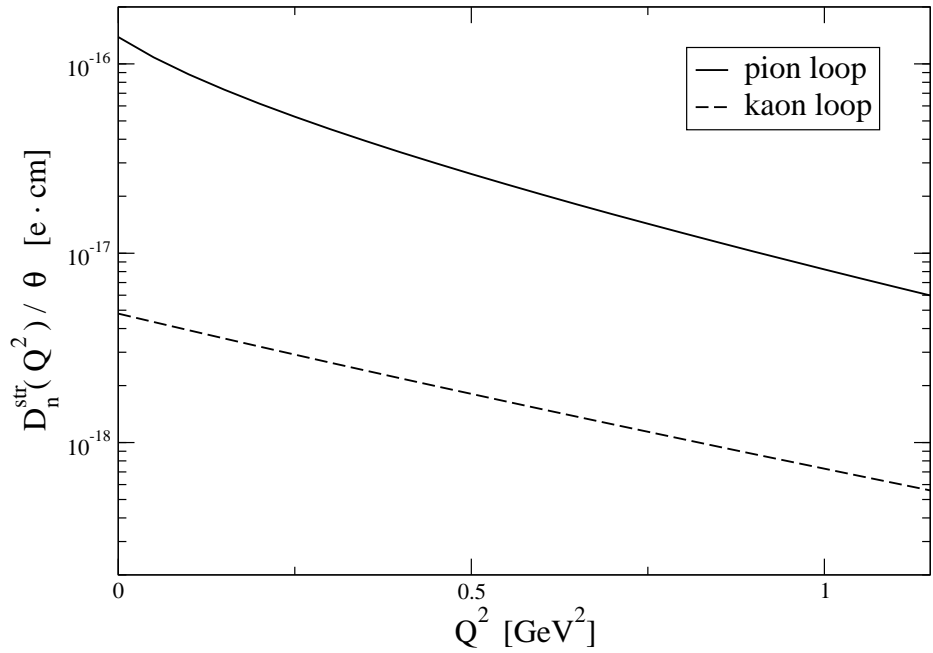}
\end{center}
\caption{The contributions from pion and kaon loops to the neutron
EDM form factor induced by strong CP violating $\theta$ term.}
\end{figure}

\newpage

\begin{table}
\caption{Theoretical estimates of the neutron EDM
induced by strong CP violation in units of
$|\theta| \times 10^{-16}$ $e \cdot$cm.}

\vspace*{.5cm}
\begin{tabular}{|c|l|l|}
\hline 
 $\; \; \; \; |d_n^{str}| \; \; \; \;$ & Model &
Reference \\ \hline
 $2.7$ & MIT bag model
 & \quad Baluni~\cite{Baluni:1978rf}\hspace*{1.5cm}\\
 $3.6$ & Current algebra
 & \quad Crewther {\it et al.}~\cite{Crewther:1979pi}\hspace*{1.5cm}\\
 $3.3$ & Effective chiral approach
 & \quad Pich {\it et al.}~\cite{Pich:1991fq}\hspace*{1.5cm}\\
 $6.7$ & HBChPT & \quad Borasoy~\cite{Borasoy:2000pq}\hspace*{1.5cm}\\
 $3.0$ & Chiral bag model
 &\quad  Musakhanov {\it et al.}~\cite{Musakhanov:1984qy}\hspace*{1.5cm}\\
 $1.4$ & Cloudy bag model
 &\quad  Morgan {\it et al.}~\cite{Morgan:1986yy}\hspace*{1.5cm}\\
 $1.17$ & Chiral quark-meson model
 & \quad McGovern {\it et al.}~\cite{McGovern:1992ix}\hspace*{1.5cm}\\
 $2.4$ & QCD sum rules
 & \quad Pospelov {\it et al.}~\cite{Pospelov:1999mv}\hspace*{1.5cm}\\
 $1.42$   & This work &   \\
\hline 
\end{tabular}
\end{table}

\end{document}